\documentclass[preprint]{aastex}

\usepackage{graphicx}

\shorttitle{Natal Super-Star Clusters} \shortauthors{Aversa et al.}
\long\def\symbolfootnote[#1]#2{\begingroup
\def\thefootnote{\fnsymbol{footnote}}\footnote[#1]{#2}\endgroup} 

\begin{document}

\title{VLA and ATCA Search for Natal Star Clusters in Nearby Star-Forming Galaxies}

\author{Alan G. Aversa\altaffilmark{1,2}, Kelsey
  E. Johnson\altaffilmark{3,4}, and Crystal L. Brogan} \affil{National
  Radio Astronomy Observatory, 520 Edgemont Rd, Charlottesville, VA
  22903} \email{aaversa@nrao.edu} \author{W.M. Goss} \affil{National
  Radio Astronomy Observatory, P.O. Box O, 1003 Lopezville Road,
  Socorro, NM 87801-0387} \author{D.J. Pisano\altaffilmark{5}}
\affil{WVU Department of Physics, P.O. Box 6315, Morgantown, WV 26506}

\altaffiltext{1}{Steward Observatory, 933 N Cherry Ave., Tucson AZ
  85721} \altaffiltext{2}{NSF Research Experiences for Undergraduates
  program} \altaffiltext{3}{Department of Astronomy, P.O. Box 400325,
  University of Virginia, Charlottesville, VA 22904-4325}
\altaffiltext{4}{Adjunct astronomer at the National Radio Astronomy
  Observatory} \altaffiltext{5}{Adjunct assistant astronomer at the
  National Radio Astronomy Observatory, P.O. Box 2, Rt. 28/92, Green
  Bank, WV 24944-0002}

\begin{abstract} In order to investigate the relationship between the
  local environment and the properties of natal star clusters, we
  obtained radio observations of 25 star-forming galaxies within
  20~Mpc using the Very Large Array (VLA) and the Australia Telescope
  Compact Array (ATCA). Natal star-forming regions can be identified
  by their characteristic thermal radio emission, which is manifest in
  their spectral index at centimeter wavelengths. The host galaxies in
  our sample were selected based upon their likelihood of harboring
  young star formation. In star-forming regions, the ionizing flux of
  massive embedded stars powers the dominant thermal free-free
  emission of those sources, resulting in a spectral index of $\alpha
  \gtrsim -0.2$ (where $S_\nu \propto \nu^\alpha$), which we compute.
  With the current sensitivity, we find that of the 25 galaxies in
  this sample only five have radio sources with spectral indices that
  are only consistent with a thermal origin; four have radio sources
  that are only consistent with a non-thermal origin; six have radio
  sources whose nature is ambiguous due to uncertainties in the
  spectral index; and sixteen have no detected radio sources.  For
  those sources that appear to be dominated by thermal emission, we
  infer the ionizing flux of the star clusters and the number of
  equivalent O7.5~V stars that are required to produce the observed
  radio flux densities. The most radio-luminous clusters that we
  detect have an equivalent of $\sim$$7 \times 10^3$~O7.5~V stars, and
  the smallest only have an equivalent of $\sim$$10^2$ ~O7.5~V stars;
  thus these star-forming regions span the range of large
  OB-associations to moderate ``super star clusters'' (SSCs).  With
  the current detection limits, we also place upper limits on the
  masses of clusters that could have recently formed; for a number of
  galaxies we can conclusively rule out the presence of natal clusters
  significantly more massive than the Galactic star-forming region
  W49A ($\sim 5\times10^4$~M$_\odot$).  The dearth of current massive
  cluster formation in these galaxies suggests that either their
  current star formation intensities have fallen to near or below that
  of the Milky Way and/or that the evolutionary state that gives rise
  to thermal radio emission is short-lived.

\end{abstract}


\section{Introduction}
Most stars are born in clusters or associations of some kind
\citep[e.g.][]{lada03,dewit05}. As a result, the clustered mode of
star formation plays a fundamental role in understanding star
formation in general. In the Galaxy, stars within a cluster can often
be individually resolved, and thus detailed studies of the resolved
structure and interplay between stars are possible. However, our
Galaxy only presents a narrow range of environmental conditions, and
we are compelled to study more distant objects in order to investigate
the impact of different environments on star cluster formation.

In relatively nearby starburst galaxies ($\lesssim20$~Mpc), recent
star formation activity has typically been resolved in massive star
clusters. The {\it Hubble Space Telescope} ({\it HST}) was
instrumental in the discovery of the so-called ``super star clusters''
(SSCs), which have been detected in significant numbers
\citep[e.g.][]{whitmore02}. Large samples of less massive clusters
have also been detected, typically following a power-law distribution
down to the completeness limits of the data. However, despite the
large samples of high quality optical data, the impact of the local
environment on massive star cluster formation is far from
understood. One of the primary obstacles has been the very nature of
star formation, which is obscured at optical wavelengths. Once
clusters have emerged from their birth material to be observable in
optical light, their birth environments can no longer directly be
probed.

To study the dependence of environment on star formation, it is
necessary to penetrate through the optically thick cocoon before the
system has had time to evolve and disperse its gas and dust, which
\citet{kobulnicky99} have estimated to last $\sim$15\% of the
lifetimes of the embedded cluster's stars. To find evidence of natal
clusters of massive stars, observations at wavelengths
$\gtrsim$~a~few~$\mu$m are necessary. High spatial resolution radio
observations at centimeter wavelengths are a powerful way to identify
the earliest phases of star-formation regions via their ``inverted"
spectral indices $\alpha$, where $S_{\nu} \propto \nu^{\alpha}$ and
$\alpha > -0.2$; this type of spectral energy distribution is similar
to that of HII regions, which exist on a smaller scale around
individual massive stars in our Galaxy, i.e., ultracompact HII regions
\citep{wood89}. If the larger ultra-dense HII regions (UDHIIs)
associated with natal clusters of massive stars emit sufficient
thermal bremsstrahlung radiation, we can detect the signature of this
radiation by its radio signature.

High spatial resolution radio observations have revealed a sample of
very young massive star clusters still embedded in their birth
material in a number of galaxies, including: NGC~5253
\citep{turner98}, He~2-10 \citep{kobulnicky99,jk03}, NGC~2146
\citep{tarchi00}, NGC~4214 \citep{beck00}, Haro~3 \citep{johnson04},
NGC~4449 \citep{reines08}, and SBS~0335-052 \citep{johnson09}. These
heavily enshrouded clusters contain hundreds to thousands of young
massive stars; these nascent stars create compact HII regions within
the dense environment and manifest themselves as optically thick
free-free radio sources, some of which have been confirmed as luminous
mid-infrared sources \citep{beck01,gorjian01, vacca02, reines08}.

Perhaps not surprisingly, the most massive and luminous natal clusters
were the first to be identified in nearby galaxies (as in the sample
cited above). However, if star cluster formation tends to follow a
power-law, as suggested by optical studies \citep{whitmore02}, we
should expect to find a continuum of extragalactic star clusters
ranging from objects similar to individual Galactic UCHII regions to
the massive proto-globular clusters common in starburst
galaxies. Furthermore, current theory suggests that the properties of
massive star clusters will largely be dependent on the pressure of
their formation environment \citep{elmegreen97}. Therefore, the most
vigorous starbursts host the most massive star clusters, while
relatively quiescent galaxies (like our own Milky Way) will tend to
contain only low mass clusters and associations. If we wish to
understand how massive star cluster formation depends on the local
environmental properties as well as to understand it in a statistical
sense, we must fill in the continuum between galactic UCHIIs and natal
SSCs with the aim of building a large sample. For this purpose, the
observations presented here are part of an effort to increase the
known sample of natal clusters in relatively nearby galaxies. We use
the NRAO\footnote{The National Radio Astronomy Observatory is a
  facility of the National Science Foundation operated under
  cooperative agreement by Associated Universities, Inc.} Very Large
Array (VLA) and the ATNF\footnote{The Australia Telescope National
  Facility, a division of the Commonwealth Scientific and Industrial
  Research Organisation, operates ATCA.} Australia Telescope Compact
Array (ATCA) to image 25 galaxies selected for their likelihood of
containing natal star formation.

\section{Observations}
\subsection{The Sample}
This sample includes 25 galaxies that were selected for this study
based on their distance and their likelihood of containing natal star
formation. Indicators of possible natal star formation included either
(1) membership in the Markarian UV catalog, the Arp catalog of
irregular and tidally interacting galaxies, or the VV catalog of
interacting galaxies (18 out of the 25 galaxies),
or (2) evidence of Wolf-Rayet (W-R) features in the host galaxy's
spectra (10 out of the 25 galaxies), necessitating the presence of
young massive stars, or (3) identification of the galaxy as a blue
compact dwarf (BCD), again an indicator of recent star formation (6
out of the 25 galaxies). Several of the galaxies in this sample fall
into more than one of these three categories.  Due to sensitivity and
spatial resolution limitations, only galaxies within $\sim$20~Mpc were
included. We give a brief overview of the selected galaxies in
\S\ref{VLA_gals}-\ref{ATCA_gals}, and their characteristics are
summarized in Table~\ref{galaxycharstable}.

\subsection{VLA and ATCA Observations}
We observed the eight southern hemisphere galaxies in our sample on
2002 March in 3~cm (8.6~GHz) and 6~cm (4.8~GHz) bands with the
Australia Telescope Compact Array (ATCA) in the 6A configuration. With
the Very Large Array (VLA), we observed the 19 northern hemisphere
galaxies on 2002 February 15 and 2002 October 21 using the
A-configuration for 3.6~cm (8.5~GHz) and the C-configuration for
1.3~cm (22~GHz) observations.

For the ATCA data, we excluded visibilities with $uv$ values
$<10$~k$\lambda$ in the 6~cm band to better match the largest spatial
scale to which the 3~cm observations are sensitive.  We then imaged
the ATCA 3~cm data with a robustness parameter of 3.0 (close to
natural weighting), which helped to mitigate noise levels in that
band, and imaged the 6~cm data with a robustness of 0.0 to add more
weight to longer baselines, thereby increasing the spatial resolution
of the resulting images.  Finally, we matched the convolution kernel of
the ATCA 3~cm images to that of the 6~cm images during the imaging
process.

We followed a similar procedure to create images of the VLA data. Since
the VLA was in the A-configuration for the 3.6~cm observations, the
resulting images have higher spatial resolution than images made with
1.3~cm observations in the C-configuration. Consequently, we imaged
the 1.3~cm data with a robustness parameter of 3.0 and excluded all
visibilities $<12$~k$\lambda$ in the {\it uv}~plane. This process
approximately matches the largest spatial scale to which the data are
sensitive. The 3.6~cm data were imaged with a robustness parameter of
0.0 in order to obtain slightly higher angular resolution than with
natural weighting.

We created images of all ATCA and VLA data with the IMAGR task of the
Astronomical Image Processing System (AIPS). The parameters for the
imaging process are summarized in Table~\ref{sourcetable}, and the
resulting radio contours for galaxies with detected emission are shown
overlaid on optical or infrared images in Figures~1 to 9.

The nature of a radio source can be constrained using its spectral
index $\alpha$, where $S_\nu\propto\nu^\alpha$. Supernova remnants
typically have radio spectral indices of $\lesssim -0.2$
\citep[e.g.][]{green84, weiler86}. Thermal sources (e.g. HII regions)
can be identified by their signature thermal bremsstrahlung
emission. Purely optically thin thermal emission has $\alpha\sim-0.1$,
while in the optically thick limit the emission has $\alpha=+2$. The
specific spectral morphology of an HII region at radio frequencies is
due to a combination of size and density structure. The frequency at
which thermal emission transitions from optically thick ($\alpha=+2$)
to thin ($\alpha=-0.1$) is higher for denser HII regions.

In order to compute a spectral index 
\begin{equation}
\label{spec_index}
\alpha_{AB}=\frac{\log_{10}{(F_A/F_B)}}{\log_{10}{(\nu_A/\nu_B)}}
\end{equation}
based on two flux densities $F_A, F_B$ and frequencies $\nu_A, \nu_B$,
it is important to match the spatial scales to which the different
frequencies are sensitive in so far as possible.  Given the nature of
interferometers, it is virtually impossible to match the synthesized
beams precisely, but steps can be taken to improve the extent to which
the synthesized beams are compatible.  To this end, when imaging the
radio observations presented in this paper, we limited the $uv$ coverage
of each data set, we varied the weighting of longer and shorter
baselines, and finally we convolved the frequencies to the same
synthesized beam. (This final step does not compensate for any missing $uv$
coverage; however, it is important to match the point response
function.)

For the purposes of this paper, we consider a source to be dominated
by thermal emission if it has a spectral index that is consistent with
$\alpha >-0.2$ within $1\sigma$ uncertainty (Table~\ref{derivedproperties}).
Likewise, we consider a source to be
dominated by non-thermal emission if it has a spectral index that is
consistent with $\alpha < -0.2$ given $1\sigma$ uncertainty. However,
in many cases it is not possible to determine unambiguously whether a
source is thermal, non-thermal, or a combination thereof for two main
reasons: (1) several of the sources presented here are clearly
extended and likely contain multiple components, and (2) the spectral
indices have significant uncertainties, and many sources that
nominally appear thermal could also be consistent (within uncertainty)
with being non-thermal and {\it vice versa}; we classify these sources
as ``ambiguous.''

\section{Results}
Of the 25 galaxies in this study, only five have radio sources that
are dominated by thermal emission, four have sources that are
dominated by non-thermal emission, six have radio sources that are
ambiguous within the uncertainty limits, and sixteen have no detected
radio sources (Figures~1 to 9; Tables~3 and 4). 
To measure the flux densities, we used the AIPS++ VIEWER
program\footnote{This capability is now available within CASA
  software.} to create identical polygonal apertures around each of
our sources at 3 and 6~cm for the ATCA data and at 1.3 and 3.6~cm for
the VLA data. By using identical convolution kernels and apertures, we
are able to maximize the accuracy of the relative photometry. Errors
in the flux densities were determined by adding in quadrature the
uncertainties due to the absolute flux calibration, variation from the
sky background, and changes in the size and shape of the aperture.
For the purposes of determining the uncertainties in the spectral
index $\alpha$, the final term in the uncertainty due to variation in
aperture is neglected as identical apertures are used at both
frequencies.
See Tables~\ref{fluxtable_vla} and
\ref{fluxtable_atca} for integrated and peak flux densities as well as
spectral indices of each source.

\subsection{Ionizing Luminosities and Cluster
  Masses} \label{ionlumsec} Massive, short-lived stars drive the
thermal free-free emission we observe at radio wavelengths; hence, an
understanding of the photo-ionization rate for each of our candidate
UDHII regions enables us to predict the number of massive stars in a
cluster. Lyman continuum photons ionize these HII regions; thus, with
knowledge of the radio luminosities, we can predict a lower bound on
the Lyman continuum flux \citep{condon92}, \begin{equation}
  Q_\mathrm{Lyc}\ge 6.3\times 10^{52}\,\mathrm{ s}^{-1}
  \left(\frac{T_e}{10^4\,\mathrm{
        K}}\right)^{-0.45}\left(\frac{\nu}{\mathrm{GHz}}\right)^{0.1}
  \left(\frac{L_\mathrm{thermal}}{10^{27}\,\mathrm{erg}\,\mathrm{s}^{-1}\,\mathrm{
        Hz}^{-1}}\right). \end{equation}

A number of possible issues must be kept in mind when interpreting the
$Q_\mathrm{Lyc}$ values and radio flux densities. First, the
application of this equation assumes the emission is purely thermal
and optically thin. Contamination from non-thermal emission within the
synthesized beams is also a possible issue at the spatial resolutions
used here, and would inflate the resulting $Q_\mathrm{Lyc}$ values. To
partially mitigate these issues, it is advantageous to use flux
densities obtained at the highest radio frequency available for two
reasons: (1) the higher frequency emission suffers from less
self-absorption and is therefore more likely to be optically thin, and
(2) the higher the frequency, the less likely it is to contain a
significant amount of non-thermal contaminating flux. Second, an
electron temperature must also be assumed, and we adopt a ``typical''
HII region temperature of $T_e = 10^4$~K; the uncertainty in
$Q_\mathrm{Lyc}$ due to this assumption is $\lesssim 20$\%. Finally,
the actual $Q_\mathrm{Lyc}$ values could be higher than observed if a
significant fraction of the ionization radiation is either absorbed by
dust within the HII region or suffers from significant leakage through
a porous ISM. Thus, when these conditions are met, the
$Q_\mathrm{Lyc}$ values quoted here should be interpreted as lower
limits.

The inferred values for $Q_\mathrm{Lyc}$ of each of
the identified thermal sources are shown in
Table~\ref{derivedproperties}.  Assuming each of these thermal sources contains
an embedded star cluster, we also estimate the number of O7.5~V stars, each
having a Lyman continuum flux $Q_\mathrm{Lyc}=1.0\times
10^{49}$~s$^{-1}$ \citep{vacca94}, required to produce the observed
free-free, thermal radio flux (see Table~\ref{derivedproperties}).

We used the stellar synthesis code Starburst99 \citep{leitherer99} in
combination with $Q_\mathrm{Lyc}$ to estimate the total stellar masses
of the natal clusters. Following \citet{johnson03}, we assume each
cluster has a metallicity $Z=0.04$ and a Salpeter IMF from 1 to
100~M$_\sun$. Assuming the $Q_\mathrm{Lyc}$ values scale directly with
the cluster mass and that the clusters are $\lesssim 3$~Myr old, we
infer stellar masses for the radio detected star-forming regions
ranging between $\sim 10^4$~M$_\sun$ and $10^6$~M$_\sun$ (see
Table~\ref{derivedproperties}). The most massive of these star-forming
regions also appear to be slightly spatially extended at the
resolution of these observations and likely include a number of star
clusters that are not resolved.

\subsection{Inferred Sizes of Thermal Sources}
In order to determine the sizes of the detected sources in
Table~\ref{derivedproperties}, we initially fit a Gaussian profile to
each source, from which the synthesized beam was deconvolved in order
to estimate the actual source sizes. The major limitation of this
method is that it requires the source to be roughly Gaussian, which is
not true for many sources presented here, some of which appear to have
quite complex structure. Nevertheless, this method will, at a minimum,
provide information about whether or not a source is extended at the
resolution of the observation. Using this method, we infer the
physical sizes of the star forming regions based upon the distances in
Table~\ref{detectionlimits} and deconvolved angular sizes in
Tables~\ref{fluxtable_vla} and \ref{fluxtable_atca}, with resulting
sizes ranging between $\sim 20$~pc to a few $\sim100$~pc. These sizes
are much larger than one would expect for an individual cluster (a few
pc); thus we conclude that most of the radio-detected star-forming
regions presented here are likely to be groups of individual clusters,
which may well be at slighltly different evolutionary
states. Furthermore, these large inferred sizes allow for a complex
origin for the observed emission and potential non-thermal
contamination; there is ample room within these large regions for a
large number of ultracompact HII regions, evolved HII regions, SNR,
and other objects. Higher spatial resolution observations are clearly
required in order to disentangle the components of the complex sources
and facilitate more precise size measurements.

\subsection{Comparison to Cas~A and W49A}
In order to provide a comparison for the relative fluxes of
non-thermal and thermal sources, we calculate the expected flux
densities and detection thresholds for the Galactic supernova remnant
Cas~A, an UCHII region complex W49A at the distance of the galaxies in
this sample.  Cas~A is the canonical ``young'' core collapse SNR in
the galaxy, with the highest luminosity and youngest age of any such
remnant in the Milky Way \citep{baars77, fesen06}.  Likewise, W49A is
a benchmark star-forming region in the galaxy, with $\sim 30$
individual thermal radio sources and $Q_\mathrm{Lyc}~ 10^{51}$~s$^{-1}$
\citep[e.g.][]{depree00}.  The expected signal-to-noise for analogs to
Cas~A and W49A in our sample galaxies at 3.6~cm are shown in
Table~\ref{detectionlimits}. For example, if Cas~A would have been a
$\sim 1\sigma$ detection in a given galaxy, we would be unlikely to
identify an individual supernova remnant. If W49A would have been a
detection $\gtrsim 5\sigma$ in a given galaxy and yet no thermal
sources are detected, such a detection limit would suggest no current
star formation above this limit in that galaxy.

Given our detection limits for objects similar to Cas~A and W49A,
there are a few striking non-detections that merit further
investigation. For example, in the galaxy Mrk~1479
\citep[$\sim4.9-5.1$~Mpc,][]{tully88,makarova98}, Cas~A and W49A would have exhibed
detections at the level of $\sim 10\sigma$ and $6\sigma$,
respectively; yet no radio sources are detected. The lack of such
natal clusters could suggest that the current star formation rate in
Mrk~1479 is below that of the Milky Way. However, the inclusion of
Mrk~1479 in both the Markarian catalog of ultraviolet bright galaxies
and Vorontsov-Velyaminov (VV) catalog of interacting galaxies
suggests that it must have recently been undergoing vigorous star
formation.  Thus, the lack of detected thermal radio sources in this
case supports the hypothesis that the natal stage of a cluster's
evolution is extremely short.

\section{Conclusions}In this radio study of 25 galaxies selected based
upon their optical signposts of star formation, we expected to detect
a number of thermal radio sources originating from massive nascent
star clusters.  However, we only detect definite thermal-dominated
radio sources in five of the sample galaxies, with an additional six
galaxies hosting radio sources whose origins are ambiguous within
uncertainties.  Using the benchmark Galactic star-forming region W49A,
we put these non-detections in context; in several galaxies, W49A
would have been a $\gtrsim 3-5\sigma$ detection.  For example,
Mrk~1479 is a notable case of a UV galaxy, included in both the VV and
Markarian catalogs, with no evidence of thermal radio emission coming
from an object similar to W49A at a $\sim 6\sigma$ level.  For the
five galaxies with detected thermal radio sources, the inferred
properties of the nascent clusters range from large OB-associations to
moderate super star clusters---extending both above and below the
mass of W49A.

One hypothesis to explain the dearth of natal clusters in this sample
is that the timescale a star-forming region spends in a stage that is
detectable in radio is extremely brief. \citet{johnson04b} suggest
that SSCs may spend as little as $\sim0.5$ to $1$~Myr in the embedded
phase when they are detectable with thermal radio emission. Compared
to galaxy evolution timescales, the time interval for star formation
is very short.  Our results are consistent with this hypothesis,
however a more complete statistical sample is needed to better
constrain the timescales.  Such a larger and complete sample would
allow us to compare the number of clusters in different
evolutionary stages and track the emergence process.

While this study was intended to identify candidate natal star-forming
regions in nearby galaxies, follow-up efforts are imperative. In
particular, this work would benefit from higher spatial resolution
observations than those presented here. Such observations will be
critical for disentangling thermal and non-thermal sources in close
proximity. In addition, higher frequency radio observations will
provide much stronger constraints on the the spectral energy
distributions of the thermal radio sources.  In particular, high
spatial resolution observations at $\sim 20-50$~GHz will be useful for
better understanding the relationship between between cluster
properties and the local environment in which they are formed.

\acknowledgments A. A. would like to acknowledge the National Science
Foundation (NSF) for supporting his research through its Research
Experiences for Undergraduates (REU) program. K. E. J. gratefully
acknowledges support for this paper provided by NSF through CAREER
award 0548103 and the David and Lucile Packard Foundation through a
Packard Fellowship.

{\it Facilities:} \facility{VLA}, \facility{ATCA}

\appendix
\section{Notes on Individual Galaxies in this Sample}

\subsection{Galaxies Observed with the VLA \label{VLA_gals}}
\subsubsubsection{Arp~217 } {\bf Arp~217 (NGC~3310, VV~356/406,
  UGC~5786)}---peculiar SAB(r)bc galaxy with giant HII regions that
may have merged with another galaxy in the past
\citep{sharp96}. Situated at 19.6~Mpc with an angular size of
$3\farcm1\times2\farcm4$, it contains W-R stars in its HII clouds,
which are 12~arcsec southwest of its nucleus
\citep{zezas98}. \citet{rosa07} estimate the SFR of Arp~217 to be
7.6~M$_\sun$~yr$^{-1}$ and 9.8~M$_\sun$~yr$^{-1}$ based on H$\alpha$
and 1.4~GHz magnitudes, respectively. Using \textit{ROSAT} and
\textit{ASCA} observations, \citet{zezas98} find hard x-ray emission
in the direction of Arp~217. Based on observations at other
wavelengths, this x-ray source could not be an AGN; therefore,
\citet{zezas04}, using \textit{Chandra}, conclude that the source of
x-rays is due to star formation. \citet{elmegreen02b} find 17
superstar cluster candidates in the southern spiral arm of Arp~217.

\subsubsubsection{Arp~233} {\bf Arp~233 (Haro~2, Mrk~33, or
  UGC~05720)}---irregular, BCD galaxy of $1\farcm12\times0\farcm80$ at
$\sim$22~Mpc. In their optical survey of BCD galaxies,
\citet{gildepaz03} deduce that because Arp~233 has a \textit{B}-band
absolute magnitude $M_B<-18.15$ and a \textit{K}-band absolute
magnitude $M_K>-21$, it must be experiencing a massive
starburst. \citet{summers01} estimate, based on evolutionary synthesis
models and $M_B$, that the age of the Arp~233 starburst is 5.8~Myr and
that its mass is $6.9\times10^6$~M$_\sun$.

\subsubsubsection{Mrk~35} {\bf Mrk~35 (NGC~3353, UGC~5860,
  Haro~3)}---a BCD galaxy of $1\farcm20\times0\farcm83$. At
$\sim$13.1~Mpc it has an optical diameter of $3.8$~kpc
\citep{steel96}. Since W-R stars have been found in Mrk~35, it must
have undergone recent star formation with a peak starburst event
occurring within the last $\sim$3-6~Myr \citep{johnson04}.

\subsubsubsection{NGC~4490} {\bf NGC~4490 (UGC~7651, Arp269,
  VV~30a)}---peculiar SB(s)d galaxy at $\sim$8.4~Mpc which has
interacted with NGC~4485, northwest of NGC~4490. \citet{elmegreen98}
have determined, with {\it N}-body simulations and an analysis of
tidal morphologies, that the two interacting galaxies collided
$4\times10^8$~yr ago, about the same time the youngest star forming
regions in the galaxy pair formed. \citet{clemens98} observed with the
VLA a large H\,{\small{\sc I}} envelope surrounding the galaxies and
discuss the possibility that the H\,{\small{\sc I}} might be
primordial gas from which the NGC~4490/4485 pair formed.

\subsubsubsection{Arp~32} {\bf Arp~32 (UGC~10770, VV89)}---peculiar type SBm
galaxy pair at a distance of $\sim$17~Mpc. \citet{damjanov06}
performed multi-wavelength photometry on Arp~32 with GALEX UV, KPNO-4m
optical, near-IR, {\it Spitzer} infrared, and 20~cm VLA radio
data. They formed an SED of the galaxy pair and fit it to a galaxy
model. The best fit model to the upper companion of Arp~32 suggests
that it is an elliptical galaxy with a 5~Gyr stellar population with
10-30\% of the stars $< 1$~Gyr. For the lower component of Arp~217,
the UV SFR is 0.12~M$_\sun$~yr$^{-1}$ while the IR SFR is larger,
1~M$_\sun$~yr$^{-1}$; this suggests obscured star formation.

\subsubsubsection{Arp~263} {\bf Arp~263 (NGC~3239, UGC~5637,
  VV~95)}---peculiar IB(s)m galaxy at $\sim$9.1~Mpc. \citet{krienke90}
use H$\alpha$ and neutral hydrogen observations to detect evidence of
new star formation in Arp~263. In addition to evidence of young star
formation, they find a warped disk and tidal tails suggesting Arp~263
has interacted tidally with a currently unseen companion.

\subsubsubsection{Arp~266} {\bf Arp~266 (NGC~4861, UGC~8098, IC~3961,
  VV~797, IZW49)}---an SB(s)m starburst galaxy at $\sim$12~Mpc. Mrk~59
and I~Zw~49 are the HII regions within the galaxy. From {4650~\AA} to
{4750~\AA}, \citet{lorenzo99} find a W-R emission bump in the spectra
of the nuclear, $\sim10\arcsec\times12\arcsec$ region of Arp~266. In
their continuum map, they observe that Arp~266 has an elongated
morphology, common in W-R and merging galaxies. \citet{barth94}
observed---in H$\alpha$, H$\beta$, and [O~III]~$\lambda$5007---28 HII
regions in Arp~266. They note a correlation between the equivalent
widths of H$\beta$ emission and the excitation index
$\log{([\mathrm{O~III}]/\mathrm{H}\beta)}$.

\subsubsubsection{Arp~277} {\bf Arp~277 (VV~313)}---a galaxy pair composed of
two Im galaxies at $\sim$12~Mpc: NGC~4809 and NGC~4810. NGC~4809 is
the brighter of the two strongly-interacting
galaxies. \citet{casasola04} report that NGC~4809 has an optical
diameter of 5.0 kpc, blue luminosity of $\log{L_B=8.18}$~L$_\sun$,
dust mass of $5.4\times10^4$~M$_\sun$, H\,{\small{\sc I}} mass of
$9.3\times10^8$~M$_\sun$, and a FIR luminosity of
$\log{L_{FIR}}=8.32$~L$_\sun$. The H\,{\small{\sc I}} mass and FIR
luminosity, however, may include emission from both NGC~4809 and
NGC~4810. From optical spectroscopy of its ionized gas, NGC~4809 has
an average $T_e\sim11300$~K, an average electron density of
98~cm$^{-3}$, and an oxygen abundance of $12+\log~[O/H]= 8.21$~dex
($\sim$1/3~Z$_\sun$) \citep{kniazev04}. NGC~4810 has a blue magnitude
of $-16.1$~mag \citep{albrecht04}. Using the H$\alpha$ flux,
\citet{james04} measured the SFR of NGC~4809 as
0.25~M$_\sun$~yr$^{-1}$ and NGC~4810 as 0.13~M$_\sun$~yr$^{-1}$.

\subsubsubsection{Arp~291} {\bf Arp~291 (UGC~5832, VV~112)}---is a peculiar,
ring galaxy at $\sim$15~Mpc. It has an optical diameter of 5.5 kpc,
blue luminosity of $\log L_B=8.81$~L$_\sun$, dust mass of
$8.1\times10^4$~M$_\sun$, H\,{\small{\sc I}} mass of
$5.5\times10^8$~M$_\sun$, and a FIR luminosity of $\log
L_\mathrm{FIR}=8.50$~L$_\sun$ \citep{casasola04}. Arp~291 has an NVSS
1.4~GHz flux of 4.1~mJy and $\log L_{1.4\mathrm{
    GHz}}=20.39$~W~Hz$^{-1}$ \citep{condon02}. At this luminosity the
radio emission is most likely related to star formation and not an
AGN, which implies a star formation rate of 0.2 M$_\sun$ yr$^{-1}$
\citep{condon92}. Arp 291 is part of a group of galaxies behind the
M~96 group and has M$_\mathrm{HI}\sim3\times10^8$~M$_\sun$ and a
dynamical mass of $\sim$$2\times10^9$~M$_\sun$ \citep{schneider89}.

\subsubsubsection{Mrk~1063} {\bf Mrk~1063 (NGC 1140, VV~482)}---Seyfert type
2, dwarf, peculiar IBm galaxy at $\sim$20~Mpc. Using {\it HST}'s
Planetary Camera, \citet{hunter94} find 6--7 SSCs in the central
0.5~kpc of Mrk~1063. \citet{degrijs04} find, based on the [Fe II]
1.6~{\micron} emission line observed with Gemini South, that both the
star formation regions of Mrk~1063 have a supernova rate of $\ga
0.3$~SN~yr$^{-1}$. They find that the young massive cluster (YMC) ages
are all $\la20$~Myr.

\subsubsubsection{Mrk~1080} {\bf Mrk~1080 (NGC~1507,
  UGC~2947)}---edge-on SB(s)m galaxy at $\sim$11~Mpc. While Mrk~1080
is optically isolated with no bright companions, it does have an
H\,{\small{\sc I}}-rich companion with
$M_\mathrm{HI}=2.6\times10^7$~M$_\sun$ and a dynamical mass of
$2\times10^9$~M$_\sun$ \citep{wilcots96}. The H\,{\small{\sc I}}
distribution of Mrk~1080 also appears to be warped in the outer parts,
possibly indicative of a recent interaction.
It is an isolated galaxy with a blue luminosity of
$2\times10^9$~L$_\sun$~and $L_\mathrm{FIR}=5\times10^8$~L$_\sun$
\citep{lisenfeld07}, and is relatively gas-rich with
$M_\mathrm{HI}/L_B=0.33$~M$_\sun$~L$_\sun^{-1}$. 
Estimates of the star formation
rate range from $0.035$~M$_\sun$~yr$^{-1}$ \citep{meurer06} from
H$\alpha$ observations and $0.15$~M$_\sun$~yr$^{-1}$ calculated from
the FIR using \citep{kennicutt98} to $0.3$~M$_\sun$~yr$^{-1}$ from 1.4
GHz radio continuum \citep{condon02}. 
\citet{miller03} have detected extra-planar diffuse
H$\alpha$ emission in this galaxy, indicating that the star formation
in Mrk~1080 strongly influences its morphology. 

\subsubsubsection{Mrk~1346} {\bf Mrk~1346 (NGC~5107, UGC~8396)}---type SB(s)d
galaxy at $\sim$14~Mpc. \citet{moorsel83} detect with the Westerbork
Synthesis Radio Telescope a mass of H\,{\small{\sc I}} northwest of
the optical bulge of Mrk~1346.  \citet{leroy05} place an upper limit
on the CO in this galaxy of $\lesssim 0.87$~K~km~s$^{-1}$.
\citet{james04} report a SFR of $0.75$M$_\odot$~yr$^{-1}$.

\subsubsubsection{Mrk~1479} {\bf Mrk~1479 (NGC~5238, VV~828,
  SBS~1331+518, IZW64)}---type SAB(s)dm galaxy at $\sim$5~Mpc. This
galaxy is a BCD and part of both the Markarian and VV catalogs,
indicating that it is both UV-bright, and also shows signs of
interaction; \citet{VV77} classify it as a interacting double
system. \citet{arkhipova87} note intense H$\alpha$ emission along the
length of the galaxy, and estimate its diameter as $\sim
2$~kpc. \citet{huchtmeier88} estimate its total mass as $\sim 3 \times
10^8$~M$_\sun$.

\subsubsubsection{Mrk~86} {\bf Mrk~86 (NGC~2537, UGC~4274, Arp~6,
  VV~138)}---peculiar SB(s)m galaxy at $\sim$6~Mpc. \citet{gildepaz00}
have found three distinct stellar populations in Mrk~86, one of which
is a 30~Myr old central starburst with a mass of
$\sim9\times10^6$~M$_\sun$. They also note that there must be a global
triggering mechanism responsible for forming the at least 46 young
star formation regions.
\citet{gildepaz02} observed the $^{12}$CO $J=1-0$ and
$J=2-1$ lines of Mrk~86 and found a horseshoe-shaped distribution of
gas surrounding the galaxy's nuclear starburst.

\subsubsubsection{Mrk~370} {\bf Mrk~370 (NGC~1036, UGC~2160,
  IC~1828)}---a BCD, peculiar galaxy at $\sim$12~Mpc. \citet{cairos02}
can reproduce their observed photometry of Mrk~370 if they assume an
instantaneous starburst with a Salpeter initial mass function (IMF)
with a mass limit of 100~M$_\sun$. They find that this starburst is
$\sim3-6$~Myr old.

\subsubsubsection{Mrk~829} {\bf Mrk~829 (UGC~09560, IIZW70,
  VV~324b)}---peculiar, BCD galaxy at $\sim$18~Mpc that is interacting
with IIZW71. \citet{rosa07} estimate the SFR of Mrk~829 based on
H$\alpha$ and 1.4~GHz to be 0.2 and 0.1~M$_\sun$~yr$^{-1}$,
respectively.  \citet{kehrig08} detect HeII $\lambda$4686 emission,
indicating hard ionizing radiation related to young massive stars. 
Kehrig et al. also determine an oxygen abundance for this system of 
$12 + \rm{log(O/H)} = 7.65-8.05$.

\subsubsubsection{NGC~1156} {\bf NGC~1156 (UGC~2455, VV~531)}---galaxy
of type IB(s)m at $\sim$6~Mpc. \citet{karachentsev96}, after finding a
distance to this Magellanic-type galaxy, note that it is ``one of the
least disturbed galaxies in the Nearby Universe'' and that it is
isolated from other galaxies. Despite its seeming quiescence, NGC~1156
shows signatures of recent star formation activity: W-R emission
features and HII emission \citep{ho95}. \citet{vacca92} argue that the
number ratios of W-R- to O-type stars in galaxies showing W-R features
indicate that the galaxy's star formation must be occurring in short
bursts of timescales $\le 10^6$~yr.

\subsubsubsection{NGC~3003 } {\bf NGC~3003 (UGC~5251)}---type SBbc, W-R
galaxy. Although there is a W-R bump detected in NGC~3003
\citep{ho95}, there is also a notable lack of broad H$\alpha$
component compared to other galaxies \citep{schaerer99}.  
\citet{rossa2003} note that the ratio of 60$\mu$m to 100$\mu$m fluxes 
suggests enhanced dust temperatures due to star formation activity.  
Rossa \& Dettmar also find strong planar H$\alpha$ emission along with 
several bright H$\alpha$ emission knots.

\subsection{Galaxies Observed with the ATCA \label{ATCA_gals}}
\subsubsubsection{NGC~1313} {\bf NGC~1313 (VV~436)}---With the largest
angular extent of any of the southern hemisphere galaxies in this
sample, it spans $9\farcm1\times6\farcm9$. It is a face-on SB(s)d
galaxy at $\sim$4.2~Mpc.
NGC~1313 contains the radio-bright remnant of the Type~II supernova
SNR 1978K \citep{ryder93}. \citet{larsen04} identifies many young
stellar clusters with ground-based and \textit{HST} WFPC2 data in
NGC~1313. Based on an extensive H\,{\small{\sc I}} (1.4~GHz) map by
\citet{peters94}, the kinematics of NGC~1313 suggest that it
interacted with a dwarf galaxy that has pulled a loop of hydrogen gas
out of its plane. Far infrared (FIR) magnitudes imply NGC~1313 has an
area-normalized star formation rate (SFR)
$\Sigma_\mathrm{SFR}=4.04\times10^3$~M$_\sun$~yr$^{-1}$~kpc$^{-2}$
\citep{larsen00}, or an estimated $1.18\times10^6$~M$_\sun$~yr$^{-1}$
throughout the whole galaxy. Wolf-Rayet (W-R) features are found at
large galactocentric radii in NGC~1313 \citep{schaerer99}.

\subsubsubsection{NGC~1510} {\bf NGC~1510}---$1\farcm3\times0\farcm7$
type E0 galaxy with two central ``condensations,'' one of which
contains W-R stars \citep{schaerer99,conti91}. \citet{eichendorf84}
suggest that an interaction with the nearby NGC~1512 has triggered a
starburst in NGC~1510. \citet{storchi94} estimate the current SFR of
NGC~1510 to be 0.3~M$_\sun$~yr$^{-1}$.

\subsubsubsection{NGC~1522} {\bf NGC~1522}---$1\farcm2\times0\farcm8$
S0 peculiar galaxy at 10.6~Mpc. It has a $M_B=-16.1$~mag and a
$\log{L_{H\alpha}}=39.91$~ergs~s$^{-1}$ \citep{gildepaz03}. The source
has been detected by FUSE in the UV \citep{fox06}. Both the H$\alpha$
flux and the UV brightness point towards prolific star formation
occurring in NGC~1522. \citet{loosethuan} classify the galaxy as an iE
BCD, meaning that it has elliptical outer isophotes and irregular
inner isophotes due to star formation. \citet{malin83} note that the
outer envelope is displaced, so this galaxy shows signatures of its
interaction with NGC~1510. It is located in the NGC~1566 group of
galaxies and has an H\,{\small{\sc I}} mass of $5\times10^8$~M$_\sun$
with a $M_{HI}/L_B$ of 0.5 in solar units \citep{kilborn05}.

\subsubsubsection{NGC~3125} {\bf NGC~3125 (Tol~3)}---an irregular BCD galaxy
with an angular size of $1\farcm1\times0\farcm7$. At a distance of
11.5~Mpc, it is comprised of two bright lobes
\citep{schaerer99}. \citet{hadfield06} surveyed the W-R stars of this
galaxy and found that there are fewer than suggested by previous UV
studies. \citet{alton94} find, with an optical polarization map, that
part of NGC~3125 is a reflection nebula illuminated by a central
starburst region.

\subsubsubsection{NGC~5408} {\bf NGC~5408
  (Tol~116)}---$1\farcm6\times0\farcm8$ IB(s)m dwarf starburst galaxy
at 4.8~Mpc. Using ATCA and \textit{Chandra} observations,
\citet{soria06} detect an ultra-luminous X-ray source in NGC~5408 that
may have formed in recent starburst activity. NGC~5408, however, is
not known to contain W-R features \citep{schaerer99}.



\subsubsubsection{NGC~2101} {\bf NGC~2101}---a type IB(s)m pec galaxy
at 13.3~Mpc\footnote{From the NASA Extragalactic Database (NED)
  \url{http://nedwww.ipac.caltech.edu/}}.  \citet{hunter04} infer a
  total star formation rate of $\sim 0.2$M$_\odot$~yr$^{-1}$ or $\sim
  0.06$M$_\odot$~yr$^{-1}$kpc$^{-2}$, among the highest SFRs for any
  of the Im-type galaxies in their sample, and typical of the SFRs for
  blue compact dwarfs in their sample.

\subsubsubsection{TOL~0957-278} {\bf TOL~0957-278 (TOL 2)}---distance
of 7.1 Mpc, $M_B=-15.19$~mag, and
$\log{L_{H\alpha}}=39.86$~ergs~s$^{-1}$
\citep{gildepaz03}. \citet{rosa07} cites a slightly higher value for
$\log{L_{H\alpha}}=40.41$ and derived a star formation rate of
0.3~M$_\sun$~yr$^{-1}$. \citet{loosethuan} classify it as an
iE~BCD. It is a possibly merging HII galaxy with strong emission lines
\citep{smith76} and a possible signature of W-R stars
\citep{kunth85,conti91,vacca92,mendez00}. 
The entire galaxy contains about $10^5$~M$_\sun$ of ionizing
stars---about 540 O5V equivalent stars. The largest optical knot has a
linear size of 225~pc \citep{mendez00}. Ages of the knots, derived
from optical emission lines and broadband colors, are $\sim$5-10~Myr
\citep{mendez00}. The galaxy has an H\,{\small{\sc I}} mass of
$3\times10^8$~M$_\sun$ \citep{barnes01}. 

\begin{deluxetable}{lllcc}
\tabletypesize{\scriptsize}
\tablewidth{0pt}
\tablecaption{Observed Galaxies with the VLA and the ATCA\label{galaxycharstable}}
\tablehead{\colhead{Galaxy} & \colhead{Alternate Name} & \colhead{Classification(s)\tablenotemark{a}} & \colhead{Distance} \\
\colhead{} & \colhead{} & \colhead{} & \colhead{(Mpc)} }
\startdata
\textbf{VLA Targets}\\
Arp~217 & NGC~3310 & SAB(r)bc pec HII &14.4  \\
Arp~233 & Haro~2 & Im pec HII  &20.4 \\
Arp~263 & NGC~3239 & IB(s)m pec  &9.1 \\
Arp~266 & NGC~4861 & SB(s)m: Sbrst &11.9 \\
Arp~277 & VV313 & Mult pec   &11.8 \\
Arp~291 & UGC~05832 & Mult pec   &15.4 \\
Arp~32 & UGC~10770 & SBm pec   &17.8 \\
Mrk~1063 & NGC~1140 & IBm pec:;HII Sy2 &20.2 \\
Mrk~1080 & NGC~1507 & SB(s)m pec?  &11.1 \\
Mrk~1346 & NGC~5107 & SB(s)d? sp  &13.8 \\
Mrk~1479 & NGC~5238 & SAB(s)dm   &4.9 \\
Mrk~35 & NGC~3353 & BCD/Irr HII  &13.8 \\
Mrk~370 & NGC~1036 & Pec?    &11.7 \\
Mrk~829 & UGC~09560 & pec; BCDG HII &17.5 \\
Mrk~86 & NGC~2537 & SB(s)m pec  &6.14 \\
NGC~1156 & UGC~02455 & IB(s)m  &6.11 \\
NGC~3003 & UGC~05251 & SBbc   &19.8 \\
NGC~4490 & ARP~269 & SB(s)d pec  &8.36 \\
\hline
\textbf{ATCA Targets}\\
NGC~1313 & ESO~082-~G~011 & SB(s)d HII   & 4.19 \\
NGC~1510 & ESO~250-~G~003 & SA0 0 pec?;HIIBCDG & 10.4 \\
NGC~1522 & ESO~156-~G~038 & (R')S0 0 : pec  & 10.0 \\
NGC~2101 & ESO~205-~G~001 & IB(s)m pec   & 13.7 \\
NGC~3125 & ESO~435-~G~041 & S BCDG    & 12.3 \\
NGC~5408 & ESO~325-~G?047 & IB(s)m HII   & 5.01 \\
TOL~0957-278 & ESO~435-IG~020 & Merger? HII  & 10.4 \\
\enddata
\tablenotetext{a}{From the NASA Extragalactic Database (NED) \url{http://nedwww.ipac.caltech.edu/}}
\end{deluxetable}

\begin{deluxetable}{lcccc}
\tabletypesize{\scriptsize}
\tablewidth{0pt}
\tablecaption{Imaging Parameters of the Observed Galaxies\label{sourcetable}}
\tablehead{\colhead{Galaxy} & \colhead{Band}
& \colhead{Convolved Beam Size\tablenotemark{a}} & \colhead{Position Angle} & \colhead{Noise} \\
\colhead{} & \colhead{(cm)} & \colhead{(arcsec)} & \colhead{(deg)} & \colhead{($\mu$Jy bm$^{-1}$)} }

\tablenotetext{a}{Applied via convolution in the imaging process}

\startdata
\textbf{VLA Targets}\\
Arp 217     & 3.6 & $1.0\times0.8$ & -81.7 & 26\\
       & 1.3 & $1.0\times0.8$ & -81.7 & 39\\
Arp 233     & 3.6 & $1.0\times0.8$ & -78.2 & 24\\
       & 1.3 & $1.0\times0.8$ & -78.2 & 37\\
Arp 263\tablenotemark{b} & 3.6 & $0.4\times0.3$ & -57.8 & 24\\
Arp 266\tablenotemark{b} & 3.6 & $0.3\times0.2$ & 79.9 & 27\\
Arp 277\tablenotemark{b} & 3.6 & $0.3\times0.3$ & 40.2 & 26\\
Arp 291\tablenotemark{b} & 3.6 & $0.3\times0.3$ & -50.4 & 26\\
Arp 32\tablenotemark{b} & 3.6 & $0.3\times0.3$ & -44.8 & 24\\
GO 20127\tablenotemark{b} & 3.6 & $0.2\times0.2$ & -7.0 & 36\\
Mrk 1063\tablenotemark{b} & 3.6 & $0.4\times0.2$ & -22.1 & 22\\
Mrk 1080\tablenotemark{b} & 3.6 & $0.4\times0.2$ & -31.1 & 24\\
Mrk 1346\tablenotemark{b} & 3.6 & $0.3\times0.2$ & 86.3 & 25\\
Mrk 1479\tablenotemark{b} & 3.6 & $0.3\times0.2$ & -85.0 & 24\\
Mrk 35     & 3.6 & $0.9\times0.8$ & -85.2 & 25\\
       & 1.3 & $0.9\times0.8$ & -85.2 & 49\\
Mrk 370\tablenotemark{b} & 3.6 & $0.3\times0.2$ & -33.2 & 22\\
Mrk 829\tablenotemark{b} & 3.6 & $0.3\times0.2$ & -83.7 & 24\\
Mrk 86\tablenotemark{b} & 3.6 & $0.3\times0.2$ & -82.4 & 27\\
NGC 1156\tablenotemark{c} & 3.6 & $0.3\times0.2$ & -49.4 & 26\\
NGC 3003\tablenotemark{b} & 3.6 & $0.5\times0.3$ & -63.7 & 30\\
NGC 4490     & 3.6 & $0.8\times0.7$ & 12.6 & 55\\
       & 1.3 & $0.8\times0.7$ & 12.6 & 20\\
\hline
\textbf{ATCA Targets}\\
NGC~1313  & 3 & $2.2\times1.5$ & -5.8 & 48\\
    & 6 & $2.2\times1.5$ & -5.8 & 53\\
NGC~1510  & 3 & $2.6\times1.7$ & 1.0 & 41\\
    & 6 & $2.6\times1.7$ & 1.0 & 48\\
NGC~1522  & 3 & $2.3\times1.7$ & -3.5 & 44\\
    & 6 & $2.3\times1.7$ & -3.5 & 45\\
NGC~2101  & 3 & $2.9\times2.0$ & -5.7 & 36\\
    & 6 & $2.9\times2.0$ & -5.7 & 42\\
NGC~3125  & 3 & $3.6\times1.7$ & 0.4 & 57\\
    & 6 & $3.6\times1.7$ & 0.4 & 61\\
NGC~5408  & 3 & $2.6\times1.7$ & 0.1 & 60\\
    & 6 & $2.6\times1.7$ & 0.1 & 48\\
TOL~0957-278 & 3 & $5.4\times1.8$ & -6.3 & 68\\
    & 6 & $5.4\times1.8$ & -6.3 & 78\\
\enddata
\tablenotetext{b}{No radio sources detected in 3.6~cm observations, and thus 
1.3~cm observations were not pursued.}
\tablenotetext{c}{Only one band available}
\end{deluxetable}

\begin{deluxetable}{lcccccccccc}
\tabletypesize{\scriptsize}
\tablewidth{0pt}
\tablecaption{Observed Properties of Detected VLA Radio Sources\label{fluxtable_vla}}
\rotate
\tablehead{\colhead{Source} & \colhead{$\alpha$} & \colhead{$\delta$} & \colhead{Deconvolved Size\tablenotemark{b}} & \colhead{Physical Size} & \colhead{$F_{1.3\mathrm{cm}}$} & \colhead{Peak $F_{1.3\mathrm{cm}}$} & \colhead{$F_{3.6\mathrm{cm}}$} & \colhead{Peak $F_{3.6\mathrm{cm}}$} & \colhead{$\alpha$\tablenotemark{c}} \\
\colhead{} & \colhead{(J2000)} & \colhead{(J2000)} & \colhead{(arcsec)} & \colhead{(pc)} & \colhead{(mJy)} & \colhead{(mJy~bm$^{-1}$)} & \colhead{(mJy)} & \colhead{(mJy~bm$^{-1}$)} }
\startdata
Arp~233n\tablenotemark{a}  & 10 32 31.80 & +54 24 04.0 & $1.8\times1.2$ & $180\times120$ & $1.1\pm0.2$ & $0.3\pm0.1$ & $0.8\pm0.2$ & $0.3\pm0.1$ & $0.3\pm0.2 $ \\
Arp~233s                   & 10 32 31.97 & +54 24 02.4 & $1.9\times1.1$ & $190\times110$ & $1.7\pm0.4$ & $0.6\pm0.1$ & $1.5\pm0.5$ & $0.7\pm0.1$ & $0.1\pm0.3 $ \\
Arp~217e\tablenotemark{a}  & 10 38 44.83 & +53 30 05.0 & $2.5\times1.3$ & $180\times 90$ & $1.4\pm0.3$ & $0.6\pm0.1$ & $1.0\pm0.2$ & $0.5\pm0.1$ & $0.4\pm0.2 $ \\
Arp~217d                   & 10 38 45.87 & +53 30 12.1 & $0.5\times0.5$ & $ 40\times 40$ & $1.3\pm1.3$ & $1.0\pm0.1$ & $1.7\pm1.7$ & $1.5\pm0.1$ & ---          \\
Arp~217c                   & 10 38 46.53 & +53 30 06.4 & $0.2\times0.2$ & $ 10\times 10$ & $0.2\pm0.1$ & $0.3\pm0.1$ & $0.4\pm0.2$ & $0.6\pm0.1$ & $-0.7\pm0.4$ \\
Arp~217b                   & 10 38 46.69 & +53 30 11.8 & $1.7\times1.1$ & $120\times 80$ & $0.5\pm0.1$ & $0.4\pm0.1$ & $0.5\pm0.1$ & $0.5\pm0.1$ & $-0.0\pm0.2$ \\
Arp~217a                   & 10 38 46.93 & +53 30 16.8 & $0.9\times0.7$ & $ 60\times 50$ & $0.6\pm0.2$ & $0.5\pm0.1$ & $0.5\pm0.2$ & $0.5\pm0.1$ & $0.2\pm0.3 $ \\
Mrk~35w\tablenotemark{a}   & 10 45 21.96 & +55 57 39.8 & $1.4\times0.9$ & $ 90\times 60$ & $1.5\pm0.2$ & $0.8\pm0.1$ & $1.5\pm0.3$ & $0.9\pm0.1$ & $0.0\pm0.2 $ \\
Mrk~35e\tablenotemark{a}   & 10 45 22.02 & +55 57 40.1 & $1.2\times1.0$ & $ 80\times 70$ & $1.3\pm0.2$ & $0.8\pm0.1$ & $1.2\pm0.3$ & $0.9\pm0.3$ & $0.1\pm0.2 $ \\
NGC~4490e                  & 12 30 29.50 & +41 39 28.4 & $0.5\times0.4$ & $ 20\times 20$ & $1.0\pm0.3$ & $0.4\pm0.1$ & $1.0\pm0.2$ & $0.6\pm0.1$ & $0.0\pm0.2 $ \\
NGC~4490bw                 & 12 30 34.44 & +41 38 25.4 & $0.8\times0.4$ & $ 30\times 20$ & $1.1\pm0.3$ & $0.4\pm0.1$ & $1.3\pm0.4$ & $0.7\pm0.1$ & $-0.2\pm0.3$ \\
NGC~4490be\tablenotemark{a}& 12 30 34.50 & +41 38 26.2 & $1.7\times1.0$ & $ 70\times 40$ & $2.2\pm0.5$ & $0.5\pm0.1$ & $2.9\pm0.6$ & $0.8\pm0.1$ & $-0.3\pm0.3$ \\
NGC~4490c                  & 12 30 34.53 & +41 38 33.3 & $0.5\times0.5$ & $ 20\times 20$ & $0.5\pm0.1$ & $0.2\pm0.1$ & $0.9\pm0.1$ & $0.5\pm0.1$ & $-0.6\pm0.1$ \\
NGC~4490d                  & 12 30 34.91 & +41 39 02.5 & $1.0\times0.6$ & $ 40\times 20$ & $0.7\pm0.1$ & $0.2\pm0.1$ & $1.3\pm0.2$ & $0.4\pm0.1$ & $-0.7\pm0.1$ \\
NGC~4490a                  & 12 30 37.73 & +41 37 58.8 & $0.8\times0.6$ & $ 30\times 20$ & $1.1\pm0.2$ & $0.3\pm0.1$ & $0.8\pm0.2$ & $0.3\pm0.1$ & $0.2\pm0.2 $
\enddata
\tablecomments{$\alpha=\log_{10}{(F_{1.3\mathrm{cm}}/F_{3.6\mathrm{cm}})}/\log_{10}{(\nu_{1.3\mathrm{cm}}/\nu_{3.6\mathrm{cm}})}$}
\tablecomments{See Table~\ref{sourcetable} for upper limits on non-detections' flux densities.}
\tablenotetext{a}{Gaussian profile did not fit source.}
\tablenotetext{b}{Sizes determined by best-fit Gaussian profile using the AIPS++ task IMAGEFITTER}
\tablenotetext{c}{Uncertainty in $\alpha$ includes uncertainty due to flux calibration and background variation, but not uncertainty due to aperture size, which is identical at the two frequencies.}
\end{deluxetable}

\begin{deluxetable}{lcccccccccc}
\tabletypesize{\scriptsize}
\tablewidth{0pt}
\tablecaption{Observed Properties of Detected ATCA Radio Sources\label{fluxtable_atca}}
\rotate
\tablehead{\colhead{Source} & \colhead{$\alpha$} & \colhead{$\delta$} & \colhead{Deconvolved Size\tablenotemark{d}} & \colhead{Physical Size} & \colhead{$F_{3\mathrm{cm}}$} & \colhead{Peak $F_{3\mathrm{cm}}$} & \colhead{$F_{6\mathrm{cm}}$} & \colhead{Peak $F_{6\mathrm{cm}}$} & \colhead{$\alpha$\tablenotemark{e}} \\
\colhead{} & \colhead{(J2000)} & \colhead{(J2000)} & \colhead{(arcsec)} & \colhead{(pc)} & \colhead{(mJy)} & \colhead{(mJy~bm$^{-1}$)} & \colhead{(mJy)} & \colhead{(mJy~bm$^{-1}$)} }
\startdata
NGC~1313snr\tablenotemark{b} & 03 17 38.66 & -66 33 03.6 & $0.4\times0.5$ & $10\times10 $  & $21\pm 12 $ & $18\pm1  $  & $27\pm16 $  & $26\pm1  $  & $-0.4\pm0.6$ \\
NGC~1313b\tablenotemark{a}   & 03 18 05.52 & -66 30 25.2 & $1.4\times1.3$ & $30\times30 $  & $1.1\pm0.3$ & $0.7\pm0.1$ & $1.1\pm0.4$ & $0.8\pm0.1$ & $0.0\pm0.4 $ \\
NGC~1313cw\tablenotemark{c}  & 03 18 37.73 & -66 29 33.7 & $1.5\times1.6$ & $30\times30 $  & $0.1\pm0.1$ & $0.2\pm0.1$ & $0.3\pm0.1$ & $0.4\pm0.1$ & $-1.6\pm0.5$ \\
NGC~1313ce\tablenotemark{c}  & 03 18 38.05 & -66 29 31.8 & $3.6\times2.2$ & $70\times50 $  & $0.3\pm0.1$ & $0.2\pm0.1$ & $1.0\pm0.2$ & $0.5\pm0.1$ & $-2.0\pm0.3$ \\
NGC~1313aw                   & 03 18 45.15 & -66 30 15.0 & $0.9\times0.6$ & $20\times10 $  & $0.5\pm0.2$ & $0.5\pm0.1$ & $0.6\pm0.2$ & $0.6\pm0.1$ & $-0.2\pm0.5$ \\
NGC~1313ae                   & 03 18 46.04 & -66 30 15.2 & $2.2\times2.4$ & $50\times50 $  & $1.8\pm0.3$ & $0.5\pm0.1$ & $2.2\pm0.3$ & $0.9\pm0.1$ & $-0.3\pm0.2$ \\
NGC~1510                     & 04 03 32.80 & -43 23 58.1 & $1.7\times1.1$ & $80\times50 $  & $0.9\pm0.6$ & $0.7\pm0.1$ & $1.0\pm0.6$ & $0.7\pm0.1$ & $-0.1\pm0.8$ \\
NGC~1522                     & 04 06 08.13 & -52 40 03.4 & $7.9\times2.8$ & $380\times140$ & $0.2\pm0.1$ & $0.1\pm0.1$ & $0.2\pm0.1$ & $0.3\pm0.1$ & $-0.0\pm0.6$ \\
NGC~3125w                    & 10 06 33.34 & -29 56 06.8 & $2.2\times1.7$ & $130\times100$ & $3.1\pm1.0$ & $2.4\pm0.1$ & $3.2\pm1.2$ & $2.5\pm0.1$ & $-0.1\pm0.4$ \\
NGC~3125e                    & 10 06 33.98 & -29 56 11.9 & $5.5\times2.9$ & $330\times170$ & $1.7\pm0.7$ & $0.9\pm0.1$ & $1.4\pm1.0$ & $1.0\pm0.1$ & $0.3\pm0.6 $ \\
NGC~5408s                    & 14 03 18.35 & -41 22 52.6 & $6.8\times2.5$ & $170\times60 $ & $3.6\pm0.8$ & $1.4\pm0.1$ & $3.8\pm0.9$ & $1.5\pm0.1$ & $-0.1\pm0.3$ \\
NGC~5408n                    & 14 03 18.67 & -41 22 50.0 & $4.5\times2.1$ & $110\times50 $ & $0.3\pm0.1$ & $0.4\pm0.1$ & $0.3\pm0.1$ & $0.4\pm0.1$ & $-0.0\pm0.3 $
\enddata
\tablecomments{$\alpha=\log_{10}{(F_{3\mathrm{cm}}/F_{6\mathrm{cm}})}/\log_{10}{(\nu_{3\mathrm{cm}}/\nu_{6\mathrm{cm}})}$}
\tablecomments{See Table~\ref{sourcetable} for upper limits on non-detections' flux densities.}
\tablenotetext{a}{\citet{larsen04} identifies this source as the star cluster n1313-341.}
\tablenotetext{b}{Known supernova remnant 1978K.}
\tablenotetext{c}{Gaussian profile did not fit source.}
\tablenotetext{d}{Sizes determined by best-fit Gaussian profile using the AIPS++ task IMAGEFITTER}
\tablenotetext{e}{Uncertainty in $\alpha$ includes uncertainty due to flux calibration and background variation, but not uncertainty due to aperture size, which is identical at the two frequencies.}
\end{deluxetable}

\begin{deluxetable}{lcccc}
\tabletypesize{\scriptsize}
\tablewidth{0pt}
\tablecaption{Inferred Properties of Thermal\tablenotemark{a} Radio Sources \label{derivedproperties}}
\tablehead{\colhead{Source} & \colhead{Size\tablenotemark{b}} & \colhead{$Q_\mathrm{Lyc}$ Lower Limit} & \colhead{Stellar Mass} &\colhead{O7.5~V Stars\tablenotemark{c}}\\

\colhead{} & \colhead{(pc)} & \colhead{($\times 10^{50}$ s$^{-1}$)} &
\colhead{($\times 10^5 M_\sun$)} &\colhead{(min. number)}}
\startdata
\textbf{VLA Targets}\\
 Arp~217a & $ 50$ & $130$ & 8 &$ 1280$ \\
 Arp~217b & $ 80$ & $110$ & 7 &$ 1070$ \\
 Arp~217e & $ 90$ & $300$ & 19 &$ 2990$ \\
 Arp~233n & $ 120$ & $470$ & 30 &$ 4720$ \\
 Arp~233s & $ 110$ & $730$ & 46 &$ 7300$ \\
 Mrk~35e & $ 70$ & $260$ & 16 &$ 2550$ \\
 Mrk~35w & $ 60$ & $300$ & 18 &$ 2950$ \\
 NGC~4490a & $ 20$ & $79 $ & 5 &$ 790$ \\
\hline
\textbf{ATCA Targets}\\
 NGC~5408n & $ 50$ & $7 $ & 0.4 &$ 70$ \\
\enddata
\tablecomments{Thermal sources tabulated here have spectral indices $\alpha>-0.2$.}
\tablenotetext{a}{In this paper sources are considered to be predominantly thermal if $\alpha -\delta \alpha > -0.2$.  Sources with $\alpha > -0.2$ but $\alpha-\delta \alpha < -0.2$ are categorized as ``uncertain.''  }
\tablenotetext{b}{This lower limit is based upon the distances in Table~\ref{detectionlimits} and deconvolved angular sizes in Tables~\ref{fluxtable_vla} and \ref{fluxtable_atca}.}
\tablenotetext{c}{Assuming O7.5~V stars have a $Q_\mathrm{Lyc} = 1.0\times10^{49}$ s$^{-1}$ \citep{vacca94}}
\end{deluxetable}

\begin{deluxetable}{lcccccc}
\tabletypesize{\scriptsize}
\tablewidth{0pt}
\tablecaption{Detection Limits of Radio Sources\label{detectionlimits}}
\tablehead{\colhead{Source} & \colhead{Distance\tablenotemark{a}} & \colhead{Cas~A
 Limit} & \colhead{``Non-thermal''} & \colhead{W49A Limit} & \colhead{``Thermal''} & \colhead{Ambiguous}\\
\colhead{} & \colhead{(Mpc)} & \colhead{($\sigma$)} & \colhead{Sources} & \colhead{($\sigma$)} & \colhead{Sources} & \colhead{Sources}}
\startdata
\textbf{VLA Targets}\\
Arp~217   & 14.4 & 0.9 & 1       & 0.5 & 3       & 1 \\
Arp~233   & 20.4 & 0.5 & \nodata       & 0.3 & 2       & \nodata \\
Arp~263   & 9.1  & 2.4 & \nodata & 1.4 & \nodata & \nodata \\
Arp~266   & 11.9 & 1.3 & \nodata & 0.7 & \nodata & \nodata\\
Arp~277   & 11.8 & 1.3 & \nodata & 0.8 & \nodata & \nodata\\
Arp~291   & 15.4 & 0.8 & \nodata & 0.5 & \nodata & \nodata\\
Arp~32    & 17.8 & 0.6 & \nodata & 0.4 & \nodata & \nodata\\
Mrk~1063  & 20.2 & 0.5 & \nodata & 0.3 & 6--7\tablenotemark{b} & \nodata\\
Mrk~1080  & 11.1 & 1.6 & \nodata & 1.0 & \nodata & \nodata\\
Mrk~1346  & 13.8 & 1.0 & \nodata & 0.6 & \nodata & \nodata\\
Mrk~1479  & 4.6  & 9.5 & \nodata & 5.6 & \nodata & \nodata\\
Mrk~35    & 13.8 & 1.0 & \nodata & 0.6 & 2       & \nodata\\
Mrk~370   & 11.7 & 1.6 & \nodata & 0.9 & \nodata & \nodata\\
Mrk~829   & 17.5 & 0.7 & \nodata & 0.4 & \nodata & \nodata\\
Mrk~86    & 6.1  & 4.7 & \nodata & 2.8 & \nodata & \nodata\\
NGC~1156  & 6.1  & 4.9 & \nodata & 2.9 & \nodata & \nodata\\
NGC~3003  & 19.8 & 0.4 & \nodata & 0.2 & \nodata & \nodata\\
NGC~4490  & 8.4  & 1.2 & 2       & 0.7 & 2       & 2 \\
\hline
\textbf{ATCA Targets}\\
NGC~1313     & 4.2  & 5.8 & 2       & 3.4 & \nodata & 4 \\
NGC~1510     & 10.4 & 1.1 & \nodata & 0.6 & \nodata & 1 \\
NGC~1522     & 10.0 & 1.1 & 1       & 0.6 & \nodata & \nodata \\
NGC~2101     & 13.7 & 0.7 & \nodata & 0.4 & \nodata & \nodata \\
NGC~3125     & 12.3 & 0.6 & \nodata & 0.3 & \nodata & 2 \\
NGC~5408     & 5.0  & 3.2 & \nodata & 1.9 & 1       & 1 \\
TOL~0957-278 & 10.4 & 0.7 & \nodata & 0.4 & \nodata & \nodata \\
\enddata
\tablecomments{We assume Cas~A at 3.6~cm (8.5~GHz) is 612~Jy at 2.8~kpc \citep{baars77}, and W49A at 3.6~cm (8.5~GHz) is 57.7~Jy at 14.1~kpc \citep{mezger67}.}
\tablenotetext{a}{Distances are galactocentric.}
\tablenotetext{b}{Although we did not detect thermal emission from Mrk~1063, we report the number of thermal sources \cite{hunter94} found.}
\end{deluxetable}


\begin{figure}
\label{arp217}
\epsscale{1}
\plotone{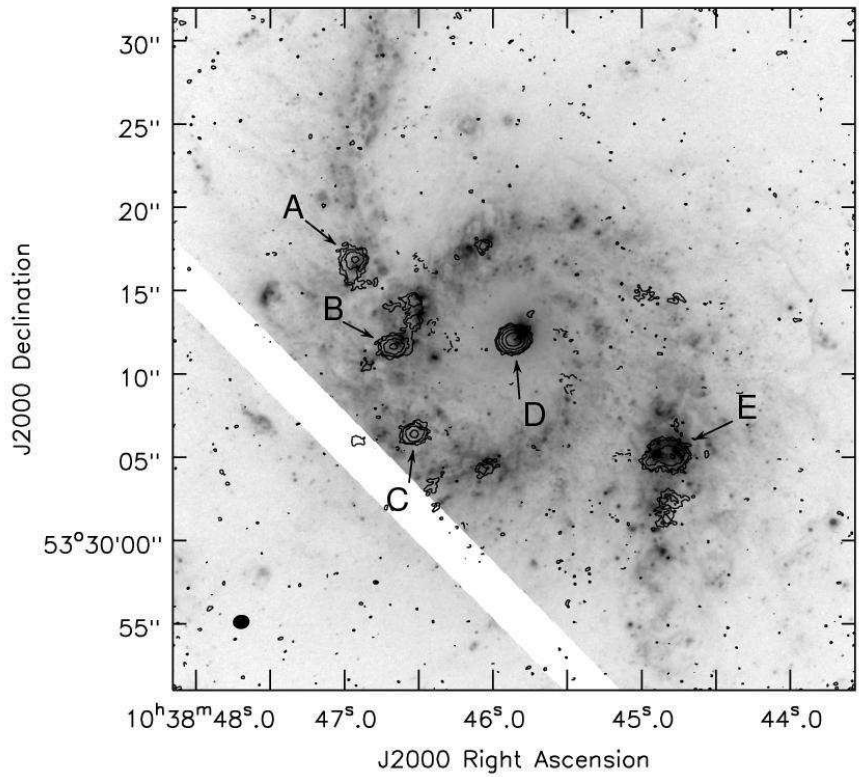}
\caption{\textbf{Arp~217}. VLA 3.6~cm (8.5~GHz) radio contours of $-3,3,5,9,17,33,57\times\sigma$ (26 $\mu$Jy bm$^{-1}$) are overlaid on an {\it HST} F658N optical image. The beam size ($1.0\times0.8$~arcsec) is indicated at the lower left. {\it HST} astronomy is precise to within $\sim1$~arcsec.}
\end{figure}

\begin{figure}
\label{arp233}
\epsscale{1}
\plotone{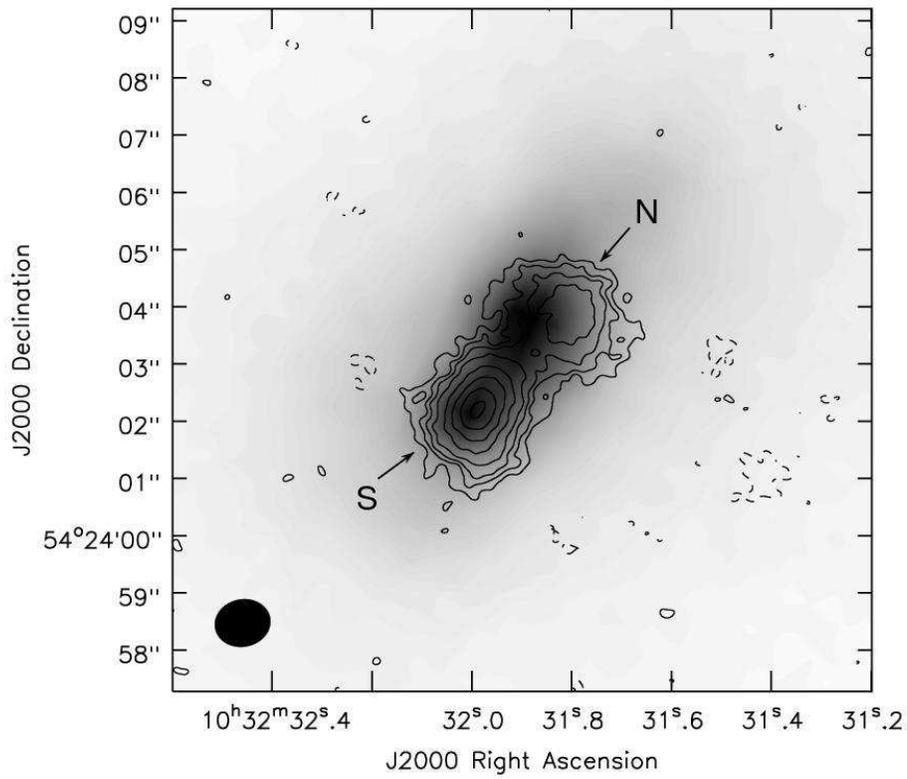}
\caption{\textbf{Arp~233}. VLA 3.6~cm (8.5~GHz) radio contours of $-3,3,5,7,10,15,20,25,29\times\sigma$ (24 $\mu$Jy bm$^{-1}$) are overlaid on this Sloan Digitized Sky Survey (SDSS) z-band optical image. The beam size ($1.0\times0.8$~arcsec) is indicated at the lower left.}
\end{figure}

\begin{figure}
\label{mrk35}
\epsscale{1}
\plotone{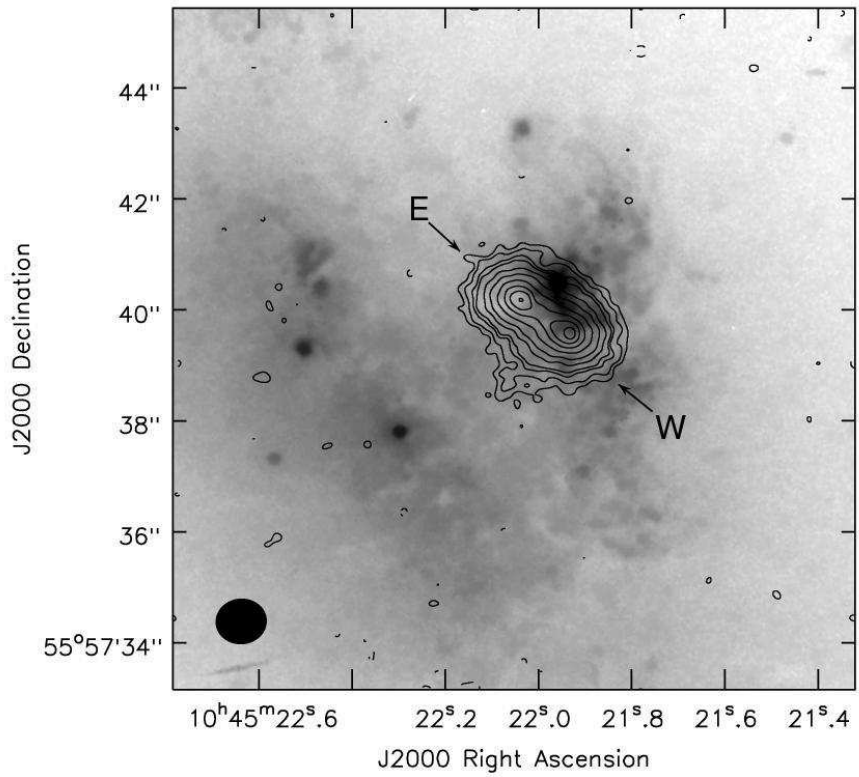}
\caption{\textbf{Mrk~35}. VLA 3.6~cm (8.5~GHz) radio contours of $-3,3,5,8,10,15,20,26,31,34,36\times\sigma$ (25 $\mu$Jy bm$^{-1}$) are overlaid on this {\it HST} F606W optical image. The beam size ($0.9\times0.8$~arcsec) is indicated at the lower left. {\it HST} astronomy is precise to within $\sim1$~arcsec.}
\end{figure}

\begin{figure}
\plotone{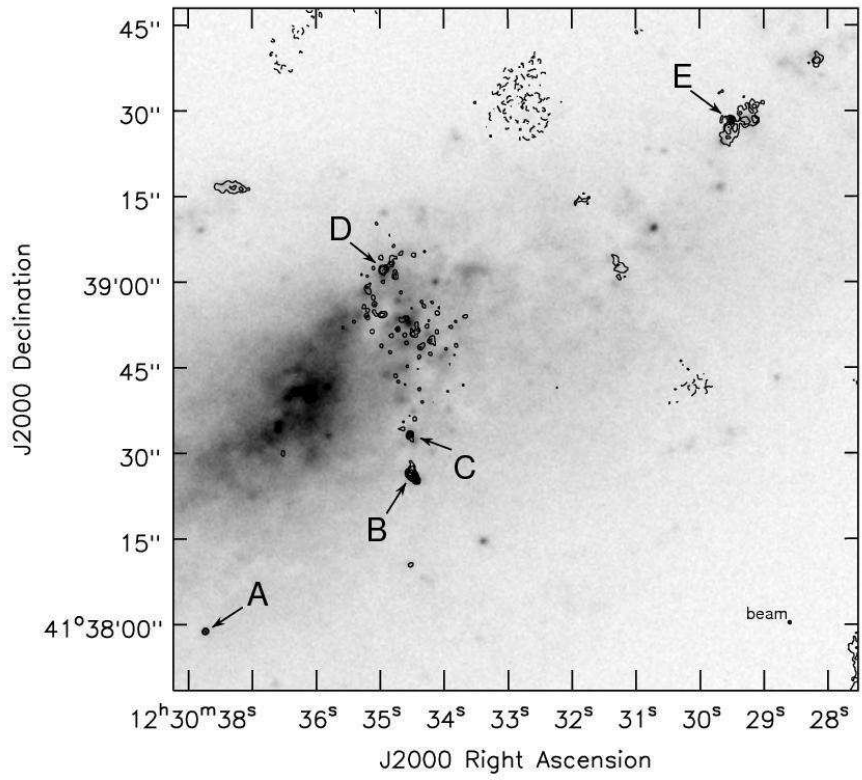}
\epsscale{1}
\caption{\textbf{NGC~4490}. VLA 3.6~cm (8.5~GHz) radio contours of $-3,3,4,6,8,10,12,14\times\sigma$ (55 $\mu$Jy bm$^{-1}$) are overlaid on this SDSS z-band optical image. The beam size ($0.8\times0.7$~arcsec is indicated at the lower left.
\label{ngc4490}
}
\end{figure}


\begin{figure}
\epsscale{1}
\plotone{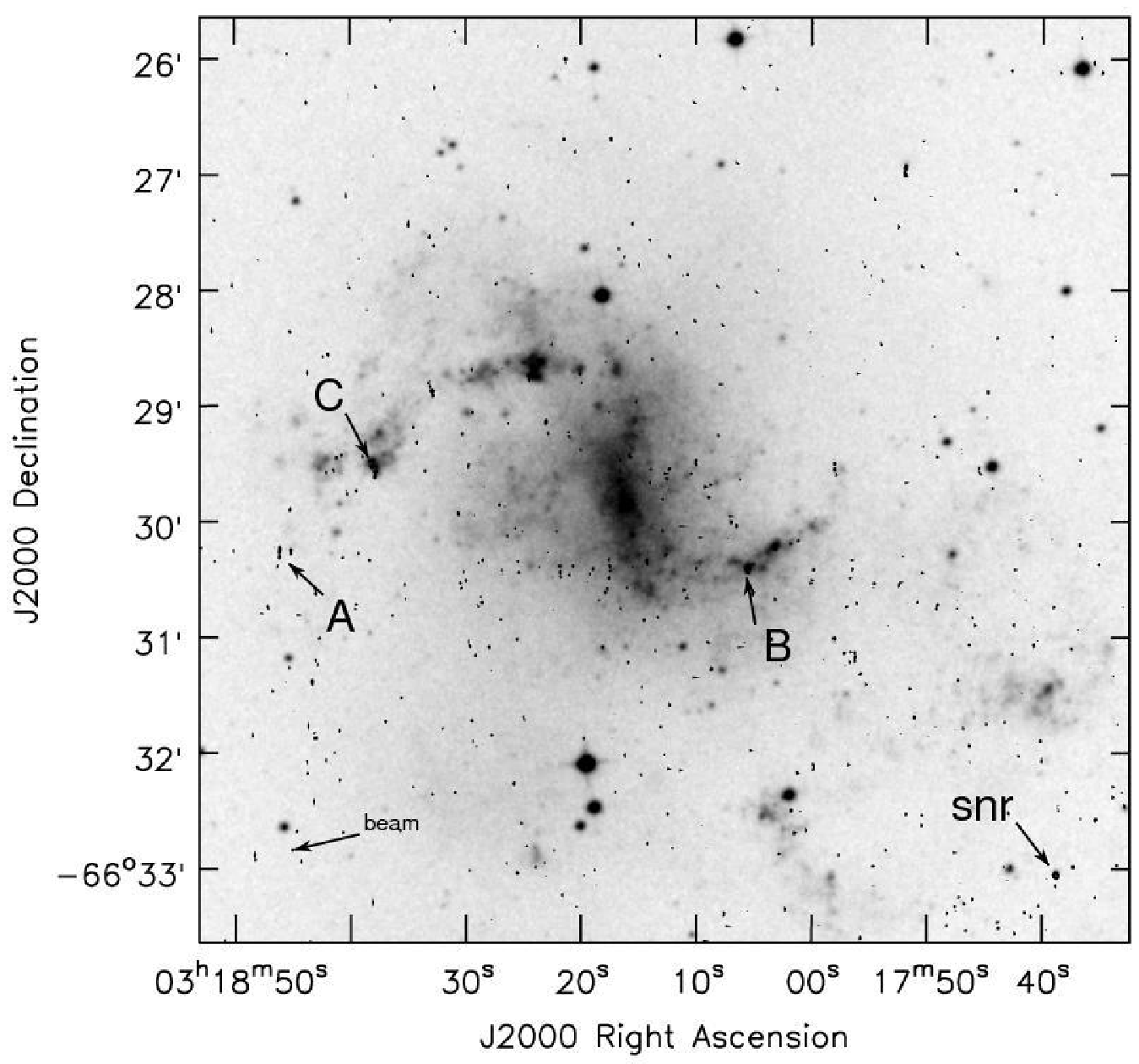}
\plottwo{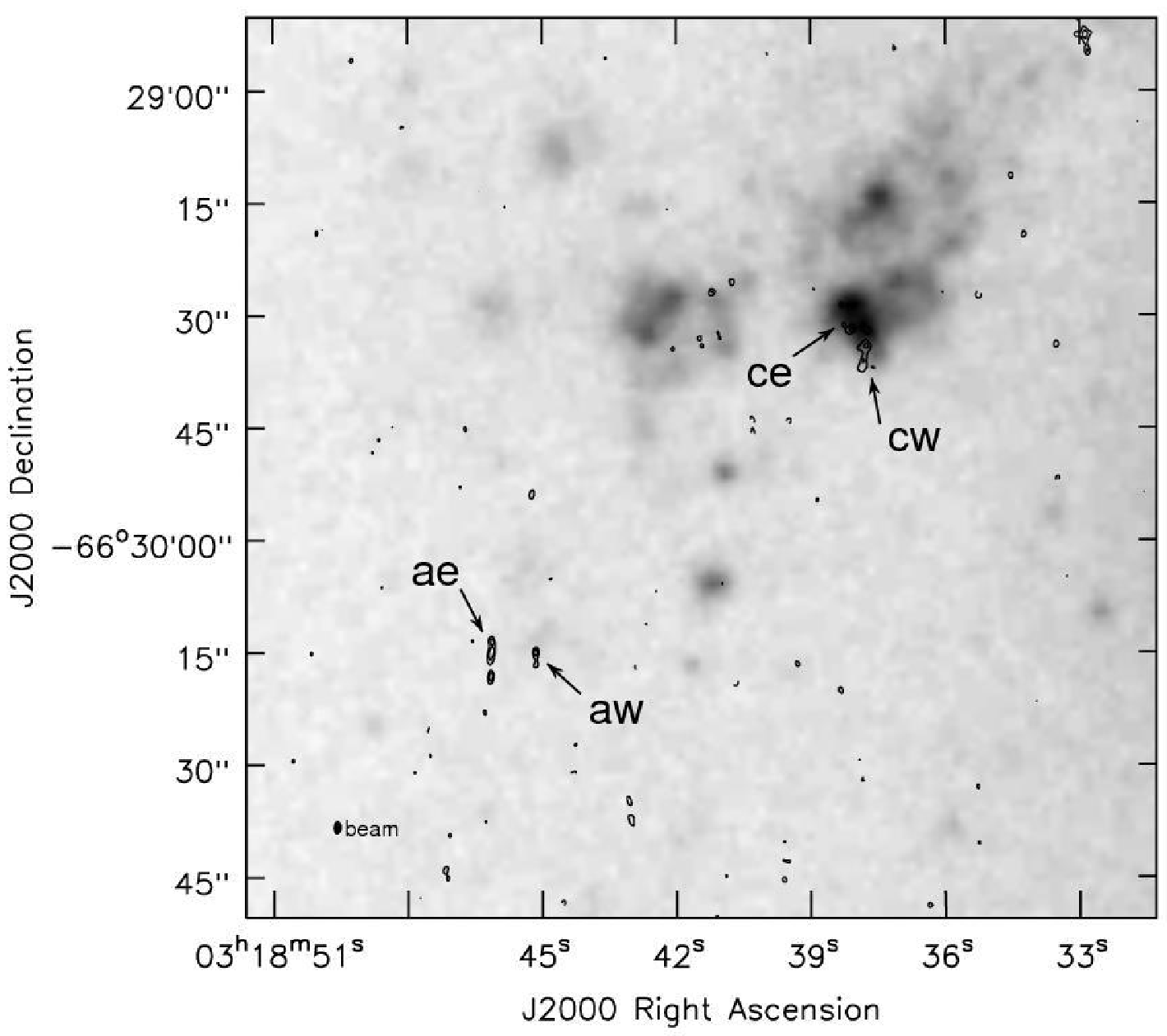}{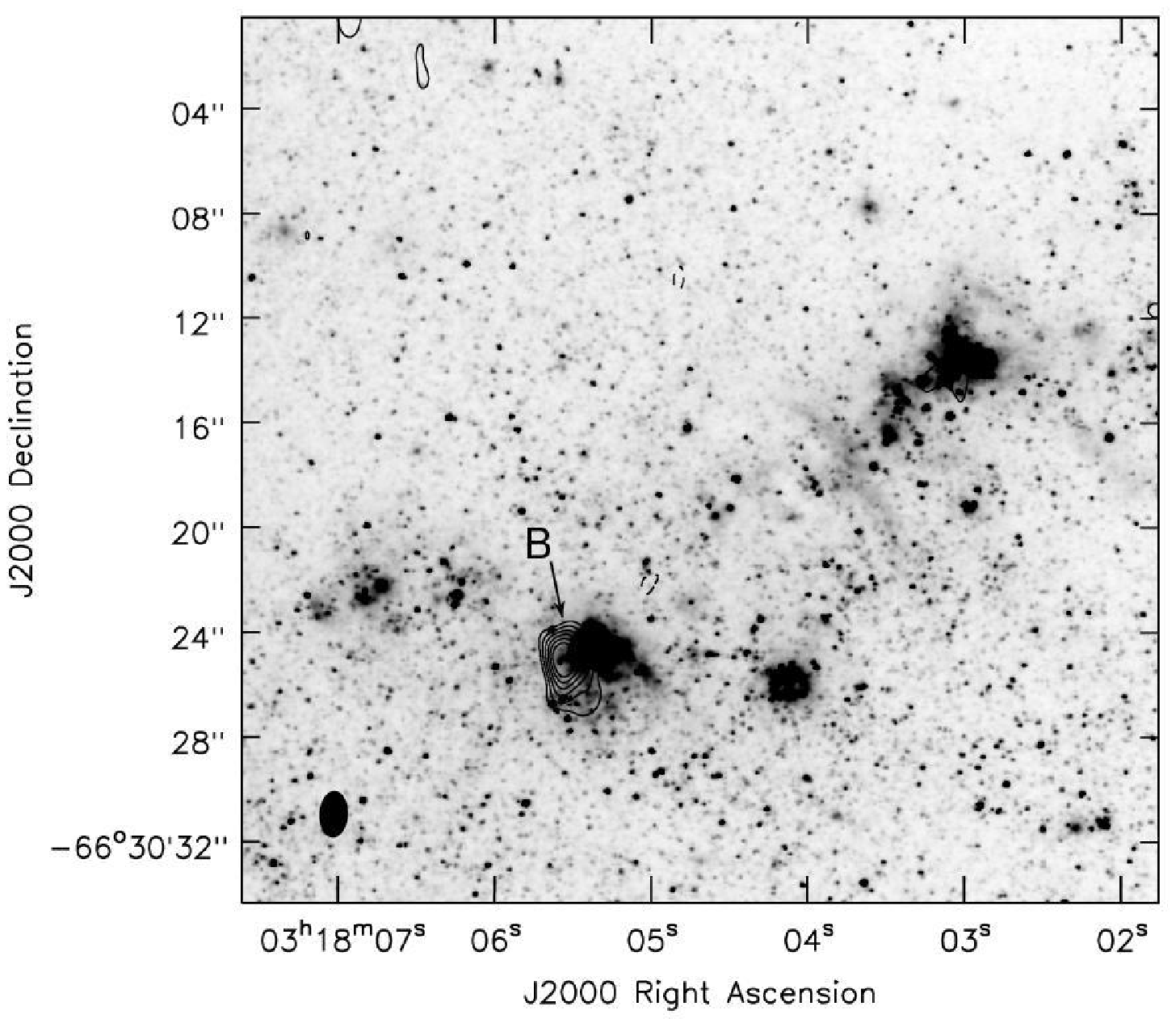}
\caption{\textbf{NGC~1313}. ATCA 6.2~cm (4.8~GHz) radio contours of $-3,3,4,5,6,7,8,10,12,14\times\sigma$ (54 $\mu$Jy bm$^{-1}$) are overlaid on a Las Campanas Observatory (LCO) 2.5~m optical image. The star formation region NGC~1313b (n1313-341 of \cite{larsen04}) is the radio-bright region on the right. SNR~1978K is also shown. The beam size ($2.5\times1.5$~arcsec) is indicated at the lower left. Blow-ups show the components of the ``A" and ``B" regions.\label{ngc1313a}
}
\end{figure}

\begin{figure}
\label{ngc1510fig}
\epsscale{1}
\plotone{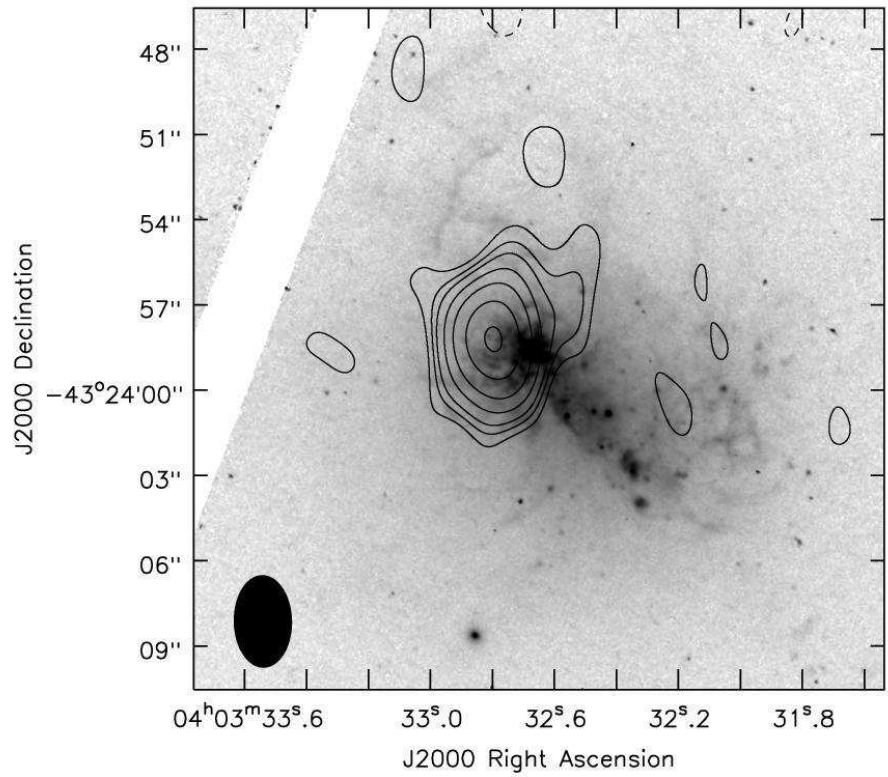}
\caption{\textbf{NGC~1510}. ATCA 3.5~cm (8.5~GHz) contours of
  $-3,3,4,5,7,10,15,21\times\sigma$ (39 $\mu$Jy bm$^{-1}$) are
  overlaid on an optical {\it HST} ACS F658N image. The beam size
  ($2.6\times1.7$~arcsec) is indicated at the lower left. {\it HST}
  astronomy is precise to within $\sim1$~arcsec.}
\end{figure}

\begin{figure}
\label{ngc1522fig}
\epsscale{1}
\plotone{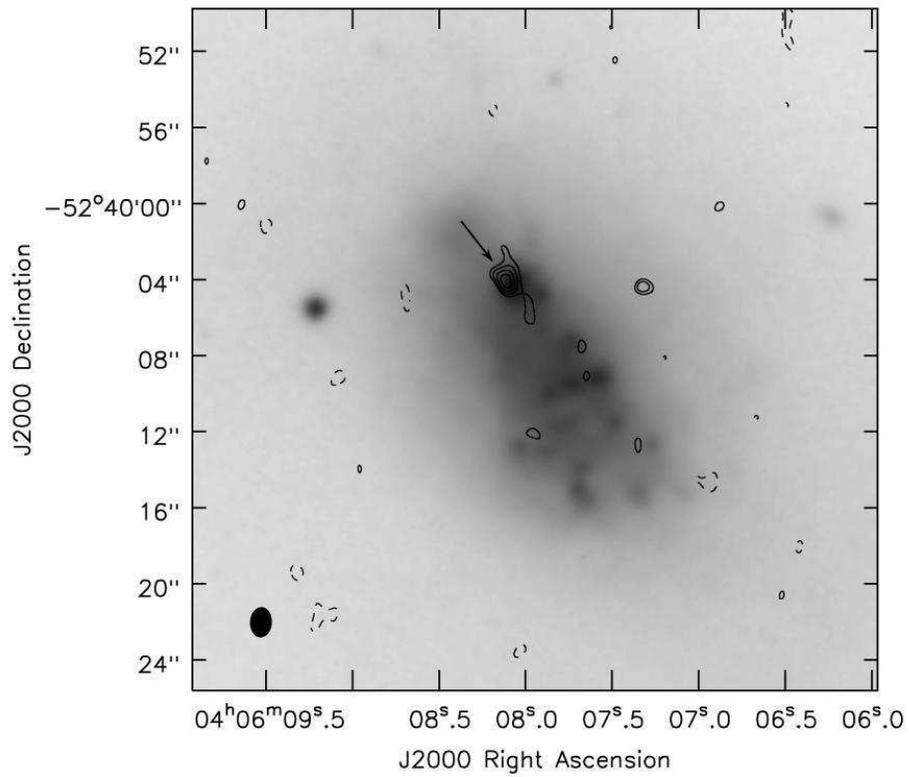}
\caption{\textbf{NGC~1522}. ATCA 3.5~cm (8.5~GHz) radio contours of $-3,3,3.5,4,4.5\times\sigma$ (39 $\mu$Jy bm$^{-1}$) are overlaid on an optical image from the LCO 2.5 m. The beam size ($2.3\times1.7$~arcsec) is indicated at the lower left.}
\end{figure}

\begin{figure}
\label{ngc3125fig}
\epsscale{1}
\plotone{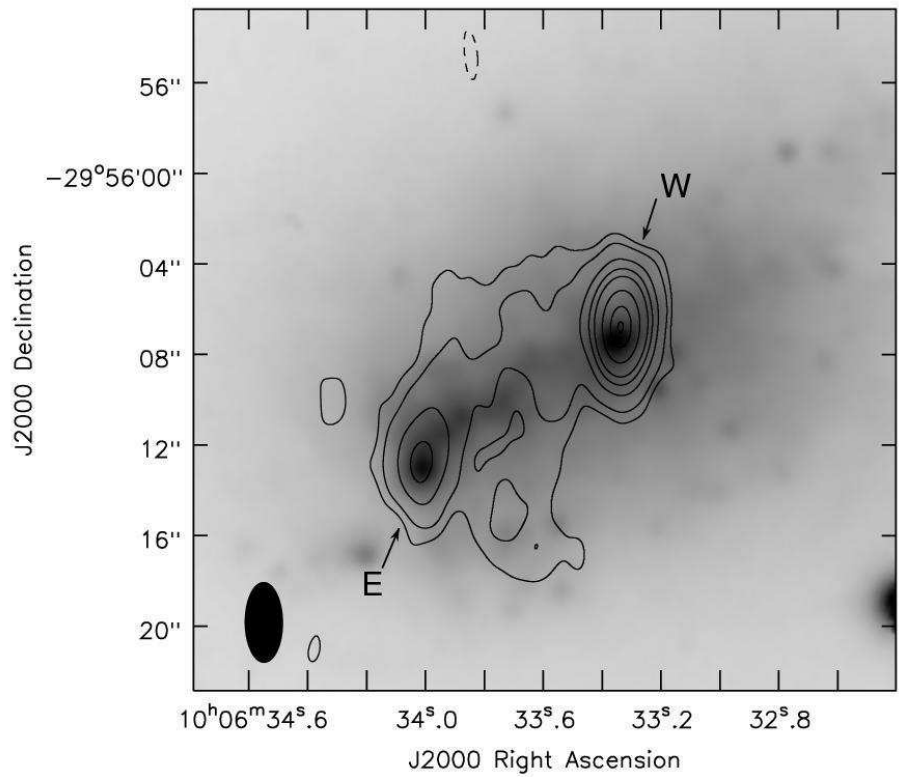}
\caption{\textbf{NGC~3125}. ATCA 3.5~cm (8.5~GHz) radio contours of $-3,3,5,10,15,20,30,40,45\times\sigma$ (53 $\mu$Jy bm$^{-1}$) are overlaid on an LCO 660~nm optical image. The beam size ($3.6\times1.7$~arcsec) is indicated at the lower left.}
\end{figure}

\begin{figure}
\label{ngc5408fig}
\epsscale{1}
\plotone{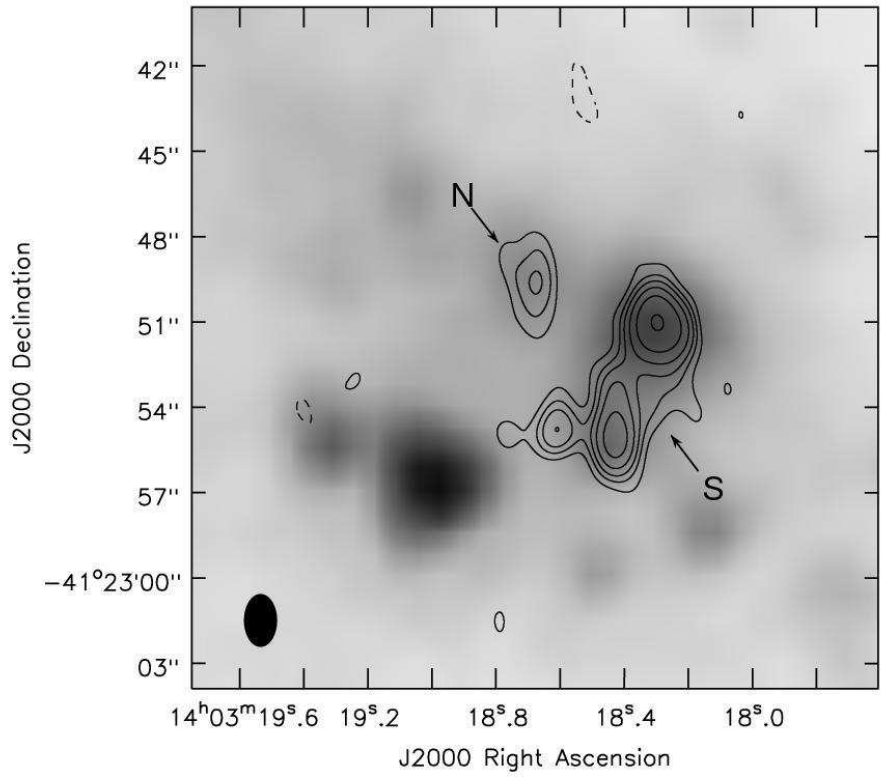}
\caption{\textbf{NGC~5408}. ATCA 3.5~cm (8.5~GHz) radio contours of $-3,3,5,7,10,15,25\times\sigma$ (42 $\mu$Jy bm$^{-1}$) are overlaid on a {\it Spitzer} IRAC 3.6~$\mu$m infrared image from the SINGS catalog \citep{kennicutt03}. The beam size ($2.6\times1.7$~arcsec) is indicated at the lower left.}
\end{figure}


\end{document}